\documentclass{aa}  
\usepackage{graphicx}
\usepackage{txfonts}
\usepackage{color}
\usepackage{hyperref}
\usepackage{listings}
\usepackage{siunitx}
\hypersetup{
    colorlinks=true,						% color of internal links
	breaklinks=true,
	linkcolor=blue,							% color of internal links
    citecolor=blue,							% color of links to bibliography
	filecolor=blue,							% color of file links
	urlcolor=black,
	unicode=false,							% non-Latin characters in Acrobat bookmarks
	pdftoolbar=true,						% show Acrobat toolbar?
	pdfmenubar=true,						% show Acrobat menu?
	pdffitwindow=false,						% window fit to page when opened
	pdfstartview={Fit},						% fits the width of the page to the window
	pdftitle={Estimating Extinction using Unsupervised Machine Learning},  %title
	pdfauthor={Stefan Meingast},			% author
	pdfsubject={PNICER},   					
	pdfcreator={Stefan Meingast},			% creator of the document
	pdfkeywords={}, 
	pdfnewwindow=true,						% links in new window
	pdfdisplaydoctitle=true					%display document title instead of file name in title b	
}

\definecolor{mygray}{rgb}{0.4, 0.4, 0.4}
\definecolor{crimson}{rgb}{0.86, 0.08, 0.24}
\definecolor{cadmiumgreen}{rgb}{0.0, 0.42, 0.24}

\definecolor{stefan}{rgb}{0.86, 0.08, 0.24}
\definecolor{joao}{rgb}{0.0, 0.42, 0.24}

\newcommand{\pnicer}{\textsc{Pnicer}}
\newcommand{\pnicers}{\textsc{Pnicer }}
\newcommand{\nices}{\textsc{Nice }}
\newcommand{\nicer}{\textsc{Nicer}}
\newcommand{\nicers}{\textsc{Nicer }}
\newcommand{\nicest}{\textsc{Nicest}}
\newcommand{\nicests}{\textsc{Nicest }}
\newcommand{\gnicer}{\textsc{Gnicer}}

\begin{document} 

\title{Estimating Extinction using Unsupervised Machine Learning}
\subtitle{}

\author{Stefan Meingast\inst{1}
        \and Marco Lombardi\inst{2}
		\and Jo\~ao Alves\inst{1}
        }
            
\institute{University of Vienna, Department of Astrophysics, T\"urkenschanzstrasse 17, 1180 Wien, Austria \\ \email{stefan.meingast@univie.ac.}
    	    \and University of Milan, Department of Physics, via Celoria 16, 20133 Milan, Italy
        }

\date{Received ...; accepted...}

\abstract{
Dust extinction is the most robust tracer of the gas distribution in the interstellar medium, but measuring extinction is limited by the systematic uncertainties involved in estimating the intrinsic colors to background stars. In this paper we present a new technique, \pnicer, that estimates intrinsic colors and extinction for individual stars using unsupervised machine learning algorithms. This new method aims to be free from any priors with respect to the column density and intrinsic color distribution. It is applicable to any combination of parameters and works in arbitrary numbers of dimensions. Furthermore, it is not restricted to color space. Extinction towards single sources is determined by fitting Gaussian Mixture Models along the extinction vector to (extinction-free) control field observations. In this way it becomes possible to describe the extinction for observed sources with probability densities, rather than a single value. \pnicers effectively eliminates known biases found in similar methods and outperforms them in cases of deep observational data where the number of background galaxies is significant, or when a large number of parameters is used to break degeneracies in the intrinsic color distributions. This new method remains computationally competitive, making it possible to correctly de-redden millions of sources within a matter of seconds. With the ever-increasing number of large-scale high-sensitivity imaging surveys, \pnicers offers a fast and reliable way to efficiently calculate extinction for arbitrary parameter combinations without prior information on source characteristics. The \pnicers software package also offers access to the well-established \nicers technique in a simple unified interface and is capable of building extinction maps including the \nicests correction for cloud substructure. \pnicers is offered to the community as an open-source software solution and is entirely written in Python.
}

\keywords{Dust, extinction -- Methods: data analysis -- Methods: statistical -- Techniques: miscellaneous}

\maketitle

% -----------------------------------------------------------------------------
% -----------------------------------------------------------------------------
\section{Introduction}
\label{sec:introduction}

Mapping the gas and dust distribution in the interstellar medium is vital to understand how diffuse clouds evolve into stars and planets, allowing for important insights on the physical mechanisms involved in processes such as cloud assemblage, evolution of dust grains, core formation and collapse, cluster formation, and the role of turbulence and feedback. Traditional techniques to map large-scale column density distributions, relying on optical star counts \citep[e.g.][]{bok1973, cambresy1999, dobashi2005} are limited to low column-densities and with the advent of near-infrared (NIR) cameras sensitive to wavelengths where clouds become transparent and reddened background stars are detected, new methods exploiting reddening have been developed to systematically study dense gas in the interstellar medium \citep[e.g.][]{lada1994,alves1998,lombardi2001,foster2008,lombardi2009,majewski2011}. The classic methods using star counts can still be applied to NIR observations \citep[e.g.][]{dobashi2011, dobashi2013} for greater dynamic range but its general limitations remain and more advanced methods making use of reddening can deliver lower-noise and more robust results for the same data set. Today, the most commonly used techniques to trace column density in the dense interstellar medium rely on 1) measuring dust thermal emission at mm and far-infrared wavelengths, 2) molecular line emission, or 3) NIR dust extinction.  

\begin{figure*}
	\centering
	\resizebox{\hsize}{!}{\includegraphics[]{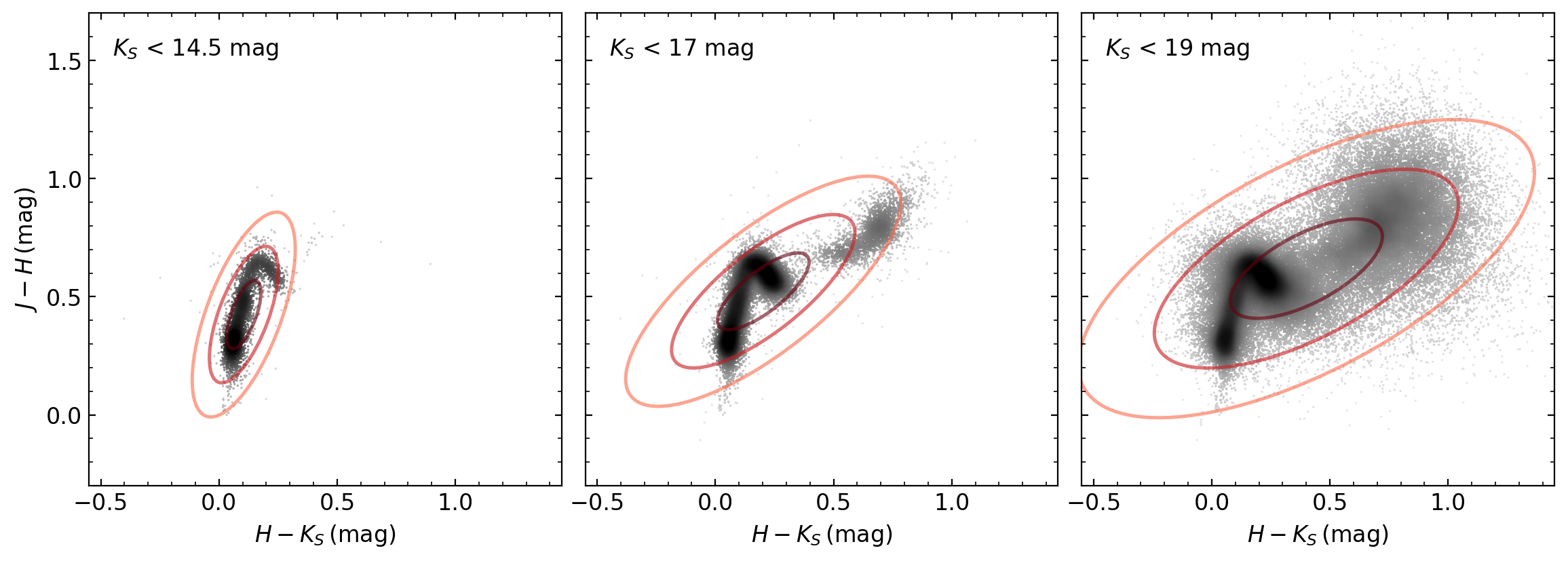}}
	\caption[]{Color-color diagram of the NIR data from \cite{meingast2016} for different magnitudes limits. The red ellipses are the covariance estimates drawn for 1, 2, and 3 standard deviations. We find a relatively small distribution width for the typical 2MASS magnitude limit at $K_S \sim 14.5$ mag in both axes which allows for an efficient description of intrinsic colors via a mean and covariance estimate. By increasing the magnitude limit, galaxies at $H-K_S \sim 0.7$, $J-H \sim 0.8$ mag significantly broaden the distribution and introduce a large statistical error when estimating intrinsic colors with a single averaged value. Also very well visible is the shift of the mean intrinsic color (i.e. the center of the ellipse) even beyond the M-branch of the stellar main sequence towards galaxies.}
    \label{img:control_covar}
\end{figure*}

Each of these techniques has its own strengths but also disadvantages. While mapping the dust thermal emission can provide large dynamic range and high-resolution maps particularly in regions away from rich stellar backgrounds, the conversion from the measured continuum emission to column-densities is far from trivial as it requires assumptions about dust emissivity and temperature. At least in regions of active star formation the temperature varies widely due to feedback processes from early-type stars. Molecular line emission can become optically thick in dense environments and furthermore relies on local (constant) conversion factors of the measured emission relative to the hydrogen abundance \citep[the so-called X-factor; for a discussion on its variations see e.g.][]{pineda2008}. The NIR dust extinction method relies on measuring color excesses of sources in the background of molecular clouds. These discrete measurements are then used to reconstruct the smooth column-density distribution (e.g. with gaussian estimators). \citet{goodman2009} showed that NIR extinction is relatively bias-free and provides more robust measurements of column density than the other tracing techniques. Ultimately, however, NIR dust extinction measurements are limited by the available number of background sources in the region of interest. In particular this method is limited in regions where even very sensitive observations will not be able to ``peer through'' high column-densities of ($A_V \gtrsim$ 100 mag) or where naturally fewer background sources are available (e.g. towards the galactic poles). Due to the declining dust opacity towards longer wavelengths one could argue that going beyond the NIR would provide further benefits but this is not the case as beyond $\sim5\, \mu$m  dust emission starts to dominate over the abrupt drop in stellar flux of background stars, acting as a bright screen, and more complex dust absorption and scattering processes play a role. Several empirical and theoretical studies have shown that there is relatively little variation in the NIR extinction law across different environments and variable dust properties \citep[][]{indebetouw2005, flaherty2007, ascenso2013}, making it ideal for robust column density measurements (and particularly for the dense gas mass distribution) in the interstellar medium.

% -----------------------------------------------------------------------------
% -----------------------------------------------------------------------------
\section{Motivation}
\label{sec:motivation}

In order to derive color excesses for individual sources it is necessary to estimate their intrinsic colors. Color excess occurs as a consequence of absorption and scattering processes when light travels through the interstellar medium and is defined via

\begin{eqnarray}
    E(m_1 - m_2)    & = & (m_1 - m_2) - (m_1 - m_2)_0 \label{eqn:excess} \\
                    & = & (m_1 - m_{1,0}) - (m_2 - m_{2,0}) = A_{m_1} - A_{m_2}
\end{eqnarray}
where the $m_i$ describe source magnitudes in different passbands (e.g. $H$ and $K_S$). The first term on the right-hand side of equation \ref{eqn:excess} refers to observed colors while $(m_1 - m_2)_0$ are intrinsic colors and the $A_{m}$ define the total extinction in the $m_i$ passband in magnitudes. Accurate estimates of intrinsic colors are not trivially derived and in principle require detailed knowledge about the characteristics of each source. For example stars need to be distinguished from (unresolved) galaxies and different stellar spectral classes show diverse intrinsic colors (e.g. main sequence and dwarf stars need to be separated from giants). The situation for main sequence stars becomes more relaxed for near, and mid-infrared wavelengths as the spectral energy distribution flattens and thus produces relatively narrow sequences in color-color space. Therefore, inferring a single average intrinsic color for all sources introduces only small statistical and systematic errors as long as this assumption does not break down. \cite{lada1994} pioneered this technique using the $H$ and $K_S$ bands to map the dust distribution throughout \object{IC 5146}. Later \cite{alves1998} improved the method and named it the \nices technique (for Near-Infrared Color Excess) and applied it in the investigation of the internal structure of the dark cloud \object{Barnard 68} based on color excess measurements made with deep NIR data \citep{alves2001}. Later this method was developed into the multi-band technique \nicers (for Near-Infrared Color Excess Revisited) by \cite{lombardi2001} which also offered an extended description of the intrinsic colors by measuring their distribution in an extinction-free nearby control field. Based on a combination of calculated covariance estimates with photometric measurement errors, color-excesses are calculated in a maximum-likelihood approach minimizing the resulting variance. Several studies of nearby giant molecular cloud complexes have used data from the Two Micron All Sky Survey \cite[2MASS, ][]{2mass} in combination with the \nicers method to study the dense gas mass distribution \cite[e.g.][]{lombardi2006, lombardi2008, lombardi2011, alves2014}. In principle, the \nicers method can be generalized and applied to any given set of color combinations as long as the interstellar reddening law at the corresponding wavelengths is well determined.

With the description of intrinsic colors via a gaussian distribution, characterized by the mean and covariance of the measured colors in an extinction-free control field, a particular problem affects applications of \nicers with very deep observations: for 2MASS data the mean $J-H$ and $H-K_S$ colors are well determined and show only a relatively small variance. For deeper and more sensitive observations, however, a large number of galaxies enters the color space. The arising issue is illustrated in Fig.~\ref{img:control_covar} where the NIR data of the control field from the \citet{meingast2016} Orion~A observations are displayed at different magnitude cuts. For these data, the completeness limit is found at $K_S \sim 19$ mag, while for 2MASS data this limit is typically found at $K_S \sim 14.5$ mag. The covariance estimates of this color combination are displayed as ellipses and are drawn for 1, 2, and 3 standard deviations. We find that the 2MASS sensitivity limits conveniently occur at magnitudes where galaxies are not detected in large quantities, resulting in a narrow distribution with standard deviations of $\sigma_{J-H} = 0.14$ and $\sigma_{H-K_S} = 0.07$ mag. Increasing the magnitude limit significantly broadens this distribution where for $K_S < 17$ mag we find $\sigma_{J-H} = 0.19$ and $\sigma_{H-K_S} = 0.16$ mag and for $K_S < 19$ mag $\sigma_{J-H} = 0.32$ and $\sigma_{H-K_S} = 0.21$ mag. For these data we therefore find that by increasing the sensitivity limit by about 5 mag, the width of the distribution in the $J-H$ vs. $H-K_S$ color space is tripled. 

The variance in the estimated extinction with \nicers, however, depends on the size of the ellipse along the extinction vector. We have tested the impact of increasing magnitude limits in the control field on the extinction error in the \nicers algorithm by creating artificial photometry without photometric errors. This ensures that the resulting errors are exclusively determined by the covariance of the control field data. For the magnitude limits of $K_S < \left\{ 14.5, 17, 19\right\}$ mag (as displayed in Fig.~\ref{img:control_covar}), we find $\sigma_{A_K} = \left\{ 0.1, 0.15, 0.2\right\}$ mag. Hence, when the covariance of the control field dominates the error budget (i.e. small photometric errors) the error in the extinction estimates with \nicers is doubled when increasing the magnitude limit from 14.5 to 19 mag. Moreover, the mean color (ellipse center) in the rightmost panel in Fig.~\ref{img:control_covar} falls between the stellar M-branch and the galaxy locus and thus the extinction, on average, will be underestimated for stars and overestimated for galaxies. We emphasize here that \nicers will still accurately reflect this behaviour by returning larger statistical errors. However, the calculated (mean) extinction estimate will be systematically shifted for both stars and galaxies.

In addition to the increased errors when dealing with deep observations, \nicers is affected by a bias when estimating color excess in highly extincted regions. In this case the populations in the science field and the control field will be different from each other as for the high column-density regions intrinsically faint sources (preferentially galaxies) will be shifted beyond the photometric sensitivity limit of the observations. For observations with a given sensitivity limit, the effect on the observed population by applying a given amount of extinction is the same as applying a magnitude cut. Looking at the $K_S < 19$ mag panel (right-most) in Fig.~\ref{img:control_covar}, one can imagine that by applying an extinction of $A_K = 2$ mag, all sources in a magnitude range from $K_S = 17$ to 19 mag will be shifted beyond the sensitivity of the survey. Thus, the observed population would be best represented by a magnitude-limited control field. In this particular example of a sensitivity limit of $K_S = 19$ mag and an extinction of $A_K = 2$ mag, the optimal control field would be limited to $K_S < 17$ mag (the middle panel of Fig.~\ref{img:control_covar}). Similarly, for an extinction of $A_K = 4.5$ mag, the intrinsic colors for the observed population in the science field should be described as given in the left-hand side panel in Fig.~\ref{img:control_covar}. \nicers in its basic implementation, however, always refers to the same (only sensitivity-limited) intrinsic color distribution, thus not optimally comparing observed populations. In other words, regardless of the amount of extinction, \nicers will always compare to intrinsic colors as given in the right-hand side panel of Fig.~\ref{img:control_covar} and therefore compare different populations. We will discuss this issue further in Sect. \ref{sec:intrinsic_colors} where we will test \nicers with a set of magnitude-limited samples.

As a solution to the problem of increasingly large uncertainties \citet[][]{foster2008} proposed to use high-resolution NIR observations to discriminate between stars and galaxies based on morphological information. They show that by including separate color excess estimates for galaxies and stars a decrease in the errors of individual pixels in extinction maps can be achieved while at the same time the resolution of the maps can be further increased. This method (\gnicer), however, comes with the handicap that prior knowledge on source characteristics is required, which oftentimes is unreliable or not available in the first place.

More recently, \citet[][]{juvela2016} present an approach to estimate extinction based on discretized intrinsic colors where the estimates are derived with Markov Chain Monte Carlo frameworks. For deep NIR observations their method delivers better results than \nicers since intrinsic colors are discretized and not described by a single parameter. Their method, however, is computationally extremely expensive and also works best when prior information about the column density distribution is available. The authors also did not investigate the possibility of including more than the three standard NIR bands $J$, $H$, and $K_S$ and only note that such an effort may come at an additional steep increase in computation time. Moreover, the method does not offer the possibility to extend the parameter space beyond photometric colors.

In this manuscript we present a new method, \pnicer\footnote{The \textit{P} in \pnicers is a reference to the calculated probability densities. We acknowledge that the original name of \nicer, \nices revisited, specifically refers to NIR photometry (despite being applicable to any other colors) and that we adopt a similar name only for consistency.}, to calculate extinction towards point sources. We characterize the extinction with a probability density function (PDF) determined by fitting Gaussian Mixture Models (GMM) along the extinction vector to extinction-free observations. Subsequently \pnicers translates the determined intrinsic parameter PDF into extinction by comparing the distribution to the observed parameters while relying on a defined extinction law. The well-established techniques to construct bias-free column-density maps from these irregularly sampled ``pencil-beam'' measurements can still be applied when the extinction is discretized via a specific metric (e.g. the expected value or the maximum of this distribution). \pnicers is exclusively data-driven, is applicable to any combination of parameter spaces (thus, not restricted to color-space like the methods above), and does not rely on any prior information on column-densities or on synthetic models. In the following discussion we will demonstrate that the method is computationally inexpensive and our implementation is capable of statistically calculating reliable intrinsic colors in multiple dimensions for tens of millions of sources in a matter of seconds. The publicly available \pnicers code is purely written in Python and is implemented in such a way as to easily allow adaptation for individual use cases.

\begin{table}
    \begin{tabular*}{\linewidth}{l @{\extracolsep{\fill}} c c c c c}
	\hline\hline
	Band                & $J$   & $H$   & $K_S$ & $W1$  & $W2$  \\
    \hline
    $\lambda$ ($\mu$m)  & 1.25  & 1.65  & 2.15  & 3.37  & 4.62  \\
    $A_{\lambda}/A_K$   & 2.5   & 1.55  & 1.0   & 0.74  & 0.54  \\
    Reference           & 1     & 1     & 1     & 2     & 2     \\
    \hline
	\end{tabular*}
    \tablebib{(1) \citet{indebetouw2005}; (2) Ascenso, J. (private communication)} 	
    \caption{Extinction law used for the verification and test setup.}
	\label{tab:extinction_law}
\end{table}

To illustrate the \pnicers design concept, its algorithms, and for subsequent verification we use data from the Vienna Survey in Orion \citep[][hereafter referred to as VISION]{meingast2016}. These data include deep NIR observations of the Orion~A molecular cloud, as well as a nearby extinction-free control field. The sources were additionally cross-correlated with the ALLWISE source catalog \citep[][]{wright2010, cutri2013} to increase the number of available parameters, but was restricted to the first and second WISE bands (hereafter referred to as $W1$ and $W2$) at 3.4 $\mu$m and 4.6 $\mu$m, respectively. For the remainder of this article we use the extinction law as given in Table \ref{tab:extinction_law}. After describing the essential functionality and design concept of \pnicers including all key algorithms in  Sect. \ref{sec:description}, Sect. \ref{sec:validation} will demonstrate the improvements and reliability by comparing the new method directly to \nicer. In addition we will also briefly discuss the software performance and availability. Section \ref{sec:summary} summarizes all key aspects of \pnicer. Information on the software structure and its dependencies along with a simplified example are given in Appendix \ref{sec:software}.

% -----------------------------------------------------------------------------
% -----------------------------------------------------------------------------
\section{Method description}
\label{sec:description}

One of the main features and strengths of \pnicers over other extinction estimators is that the method is easily applicable to any combinations and number of parameters. For instance it is easily possible to combine source colors with apparent brightness information. For this reason we will refer to individual input parameters as \textit{features}. In practice these features will mostly consist of magnitudes and colors, but in principle \pnicers can be used with any parameter as long as the effects of interstellar extinction on the given feature are known. 

\begin{figure*}[ht!]
	\centering
	\resizebox{\hsize}{!}{\includegraphics[]{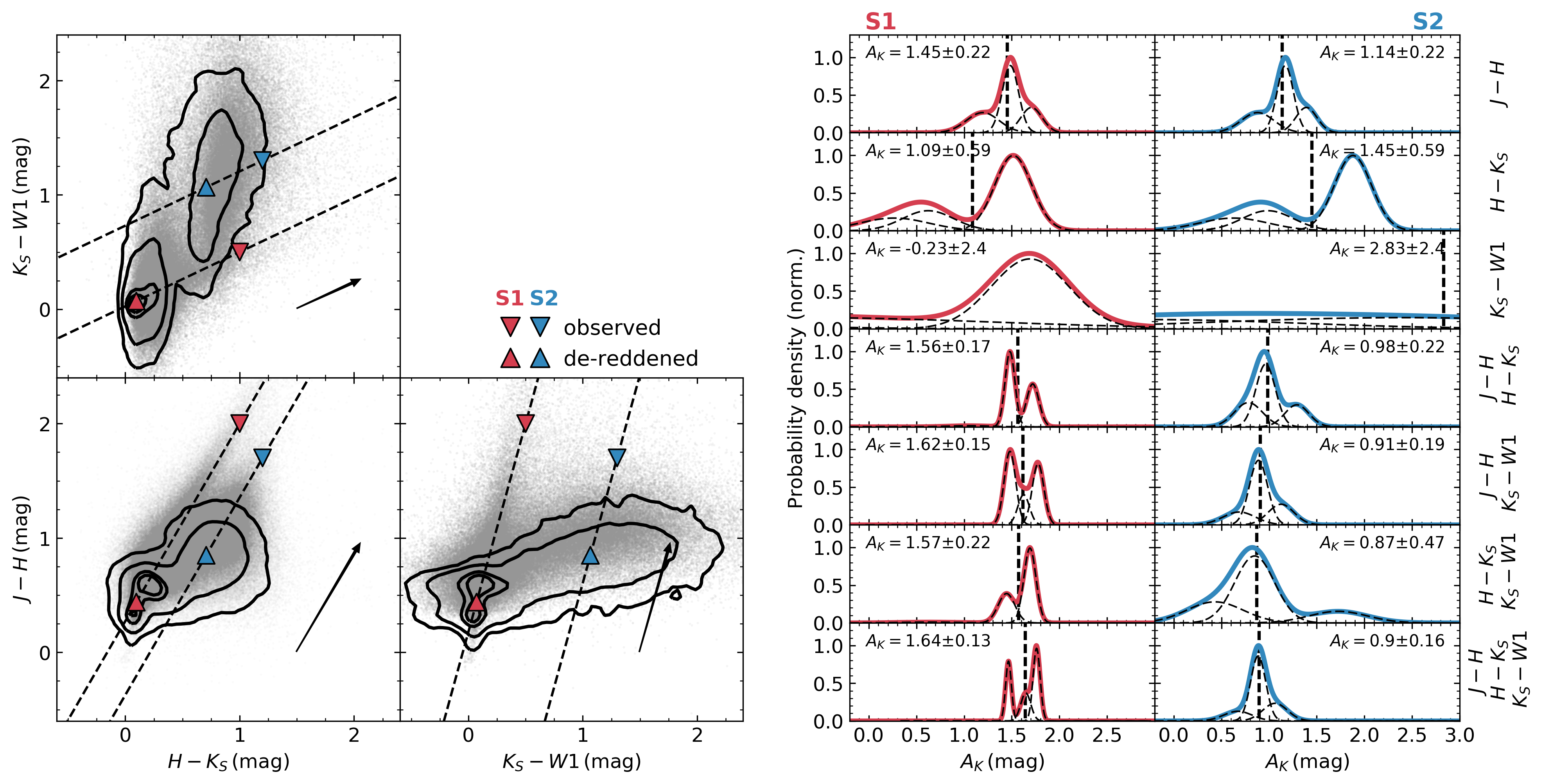}}
	\caption[]{\pnicers method concept. Left: VISION Orion~A data (gray points) and extinction-free control field number density distribution (black solid contours, levels at 0.5\%, 3\%, 25\%, and 50\% of the maximum density) for the two-dimensional feature combinations in our test setup. For two test sources S1 (red) and S2 (blue) the extinction is described with the PDF along the reddening vector (black dashed lines) in the control field. Right: Probability densities functions described by GMMs along the reddening vector for both sources (S1: red, left column, S2: blue, right column) for all possible combinations of features. Individual combinations are shown in different rows and their labels are found at the right-hand side. The individual Gaussian components are also displayed as dashed lines. The annotated extinction estimates refer to the expected value (also marked as vertical dashed lines) of the PDF and its population variance. The PDF with the minimum population variance is chosen as the best approximation for the line-of-sight extinction. Clearly some combinations offer better extinction determinations and some degeneracies are partly lifted (but not entirely resolved) for higher numbers of dimensions. In this example the extinction is estimated via the expected value of the PDFs and is marked with dashed vertical black lines.}
	\label{img:pnicer_method}
\end{figure*}

The main \pnicers algorithm can be summarized as follows: for each source for which a line-of-sight extinction is to be calculated an intrinsic feature distribution is derived along the extinction vector in the same feature space of a given extinction-free control field. The data in the control field are fitted with GMMs to construct the PDFs which serve as a probabilistic description of the intrinsic features (e.g. intrinsic colors). Due to the fact that many sources will not have measurements in all available features, \pnicers constructs all available combinations and automatically chooses the optimal (minimum variance) result. Put into machine learning terms, \pnicers uses the intrinsic feature (e.g. color) distribution from a control field to classify the intrinsic features of an extincted science field. We will now proceed to describe the details of this procedure and each individual processing step. For visual guidance, we provide a detailed three-dimensional example and follow all processing steps for two sources with artificial colors and symmetric errors in Fig.~\ref{img:pnicer_method} which will be referred to multiple times during the following discussion. 

% -----------------------------------------------------------------------------
\subsection{The multidimensional feature space}

Our example for this demonstration is limited to the four magnitudes $J$, $H$, $K_S$, and $W1$ and for illustration purposes will further be restricted to only allow the three color features $J - H$, $H - K_S$, and $K_S - W1$ with the corresponding color excesses of 0.95, 0.55, and 0.26 mag normalized to $A_K$ (see Table \ref{tab:extinction_law}). We note here that the extinction law is fixed for a single \pnicers application and the same law will be applied to all input sources (the extinction law needs to be specified upon runtime). However, in particular in the optical and perhaps also at mid-infrared wavelengths the extinction law is expected to vary with the level of extinction. It is therefore the responsibly of the user to take any such potential variations into account.

The three panels on the left-hand side of Fig.~\ref{img:pnicer_method} only show the available two-dimensional combinations of the selected color features. To calculate color excesses \pnicer, however, also uses all one-dimensional (univariate) parameter spaces, as well as all available higher dimensional combinations. In the case of our three features a total of 7 combinations is available: The univariate parameter spaces $(J-H)$, $(H-K_S)$, $(K_S-W1)$, the two-dimensional combinations $(J - H, H - K_S)$, $(J - H, K_S - W1)$, $(H - K_S, K_S - W1)$ and the three-dimensional case $(J-H, H-K_S, K_S-W1)$. Thus, the left-hand side panels of Fig.~\ref{img:pnicer_method} can be interpreted as projected views of the three-dimensional combination. If, in this case, one also allows individual passbands as features, a total of 127 combinations would be available within the seven-dimensional feature space. The practical limit of usable dimensions depends on the sampling of the control field data space: more dimensions require a proportionally larger number of sources in the control field to have a statistically well sampled feature space. In our test runs, we successfully evaluated up to nine dimensions (511 combinations). 

The solid black contours in the left panels of Fig.~\ref{img:pnicer_method} represent the number density in the control field feature space evaluated with a 0.04 mag wide Epanechnikov kernel, where the levels indicate 0.5\%, 3\%, 25\%, and 50\% of the maximum density in the given parameter space. For comparison we also show all sources from the (partly extincted) Orion~A VISION data as grey dots in the background. For this demonstration we created artificial sources which we will follow in this example. These are marked in red and blue and are denoted S1 and S2 respectively. Their ``observed'' colors are $(J - H)_{S1,S2} = \left\{2, 1.7\right\}$, $(H - K_S)_{S1,S2} = \left\{1, 1.2\right\}$, and $(K_S - W1)_{S1,S2} = \left\{0.5, 1.3\right\}$ with a symmetric error of 0.08 mag. These observed colors are marked as triangles with the tip towards the bottom. The black dashed lines are parallels to the extinction vector drawn through the original positions of S1 and S2 and the black arrow corresponds to the effect of 1 mag of extinction in the $K_S$ band. 

Already in this view it becomes apparent that estimates of intrinsic features (as described in the control field feature space) are degenerate along the extinction vector. For many sources the extinction vector will pass through different regions in the intrinsic color distribution. For example when following the extinction vector in the ($J - H$, $H - K_S$) space through the observed colors of S1 (red triangle pointing downwards), the vector passes partly through the galaxy colors and then crosses the main sequence for late type stars as well as early type stars. From this feature combination alone it is therefore not entirely clear which intrinsic colors this source should be assigned. For the ($J - H$, $K_S - W1$) combination this degeneracy with respect to the galaxy colors seems to be better resolved. 

The control field is extinction-free and it is therefore assumed to be an accurate distribution of intrinsic colors (smoothed by the photometric errors). Ideally the data for the control field should have similar completeness limits and comparable errors to accurately reflect the information in the science field. Also, stellar and galactic populations populations and number densities should be similar in the control field and the science field. These criteria are often met when dealing with data from one set of observations.

% -----------------------------------------------------------------------------
\subsection{Constructing probability density distributions}

To estimate the intrinsic feature probability distribution, \pnicers calculates the PDFs along the extinction vector (the dashed lines in the left-hand side panels of Fig.~\ref{img:pnicer_method}) in the control field feature space. Here, the number density of sources is directly interpreted as the probability distribution of intrinsic features. In order to derive this probability density along an arbitrary feature extinction vector in any number of dimensions \pnicers initially rotates the data space with the given extinction vector until only the first feature component remains non-zero. In other words, in the rotated feature space, the extinction vector has only one non-zero component and is parallel to the first feature axis. We construct the final n-dimensional rotation matrix via a sequence of applications of 

\begin{eqnarray}
	V &=& \hat{u}_1 \otimes \hat{u}_1 + \hat{u}_2 \otimes \hat{u}_2	\\
    W &=& \hat{u}_1 \otimes \hat{u}_2 - \hat{u}_2 \otimes \hat{u}_1	\\
    R &=& I_n + V\left[\mathrm{\cos}(\alpha) -1 \right] + W\sin(\alpha)
\end{eqnarray}
where $I_n$ is the identity matrix for $n$ dimensions. Here the rotation matrix $R$ allows to rotate an $n$-dimensional feature space by an angle $\alpha$ in the plane spanned by the unit vectors $\hat{u}_1$ and $\hat{u}_2$ where $\hat{u}_1 \cdot \hat{u}_2 = 0$ and $|\hat{u}_1| = |\hat{u}_2| = 1$. We apply these rotations $n-1$ times until only one component remains non-zero. The rotation of the intrinsic feature space allows to directly fit GMMs to the data on a discrete grid since the extinction vector in this space is parallel to the axis spanning the first dimension. The discretization of the grid is typically chosen to oversample the data by a factor of two with respect to the average feature errors. We found that GMMs with typically three components are sufficient to model the density distribution along the extinction vector. The total number of fitted components for the GMM defaults to three, but can be adapted by the user depending on the complexity of the feature space.

It would also be possible to derive the underlying PDF without the assumption that the distribution can be fitted with a limited number of Gaussian functions by directly calculating normalized kernel densities. However, doing so would imply to define a kernel bandwidth which may artificially broaden the distribution. In our method, all Gaussian functions in the fitted model use independent covariance matrices and thus optimally describe the underlying PDF of the intrinsic feature distribution. Furthermore, by modelling the probability density distributions with GMMs one has to store only a limited number of parameters which is particularly important when estimating extinction for large ensembles.

% -----------------------------------------------------------------------------
\subsection{Estimating extinction}

The process of creating probability density functions is repeated for all possible combinations of features. Using all combinations ensures that always the optimal feature space is selected and even a single feature can provide an extinction measurement\footnote{We only allow single features in color space, but not in magnitude space as a single magnitude can not be taken as a reliable indicator for extinction.} for cases where sources do not have measurements in all given features. The final extinction estimate described by a PDF is then chosen from the combination of features which minimizes the population variance. In almost all cases the combination with the largest dimensionality will be selected. Only when an additional feature has significantly larger errors than the other parameters, a reduced feature space may deliver better results. In addition we require at least twenty sources in the control field feature space to be present along the reddening vector, otherwise the extinction estimate would be highly biased by the low number of control field sources and the model fitting process may not converge. This case affects mostly sources with large photometric errors or ``uncommon'' observed colors such as Young Stellar Objects (YSO) which may not be well represented in the control field feature space. The process of estimating extinction is illustrated in the right-hand side panels of Fig.~\ref{img:pnicer_method} which show the extracted PDFs (in this case constructed from three independent Gaussian functions) for all seven possible combinations of our test features. These panels also display the calculated extinction when the expected value of the PDF is used (vertical dashed lines; see Sect. \ref{sec:discretization} for details). When using this estimator on just one available feature, \pnicers reproduces the \nicers results when applied to colors only. Then the expected value of the PDF is equal to the mean of the color distribution. Hence, the data points are projected onto the same intrinsic color. In reference to right-hand side panels in Fig.~\ref{img:pnicer_method} we note several characteristics here: 

\begin{itemize}

    \item[a)] Clearly, some feature combinations are better suitable than others because they show much narrower distributions. For example in the $J-H$ color alone (topmost sub-plot) all objects share a very similar color, making it ideal for extinction determinations in our case\footnote{For magnitude limited samples without galaxies $H-K_S$ shows a smaller variance compared to $J-H$.}. On the other hand, e.g. the $H - K_S$ and $K_S - W1$ colors only poorly constrains the extinction since the intrinsic feature space shows a very broad distribution. In fact, in this case we can see that all combinations which include a $J$ magnitude offer superior results.

    \item[b)] For the one-dimensional parameter spaces (top three rows) all colors share the same PDF shape for both sources (for $Ks - W1$ this is not well visible due to the extreme width of the mixture). We only observe a shift (depending on observed color) in the PDF describing the extinction.
    
    \item[c)] As expected, the best combination (i.e. smallest variance) is found when all features are available (bottom panels).
    
    \item[d)] For some combinations we observe a degeneracy in the probability density space. Consider source S1 in the $(J-H$, $H-K_S$) feature space. Moving along the reddening vector we pass the outermost edges of the galaxy locus ($J-H \sim 1$, $H-K_S \sim 0.4$) and then cross both the M sequence dwarf branch ($J-H \sim 0.6$, $H-K_S \sim 0.2$), as well as an enhancement caused by early-type stars ($J-H \sim 0.3$, $H-K_S \sim 0.1$). This is reflected by the asymmetric shape in the density profile. The double-peaked nature of this degeneracy due to the two stellar peaks is also apparent in the $(J-H$, $K_S - W1)$ feature space. The expected values for these PDFs place the intrinsic color of the source in the low probability valley between these peaks, demonstrating that such an estimator can be sub-optimal. In this example even a combination of three colors does not break the degeneracy.
    
    \item[e)] The source S2 is equally well constrained, though here, most distributions (those with $J$ band) show only one single peak clearly marking the source as a galaxy.
    
    \item[f)] In the case of estimating the extinction and error with the expected value and population variance of the distribution, we see a continuous improvement when using higher-dimensional feature spaces. This trend is expected to continue when even more features are added.
    
\end{itemize}

We again emphasize here that this technique is not limited to color space, but can be applied to any feature as long as its extinction component is known. In fact, optimal results are achieved when combining color and magnitude space since e.g. a bright source is unlikely to be a galaxy. Among all available combinations we chose the PDF that shows the smallest distribution width.

It is worth pointing out that the de-reddening process is the same for all sources and the extinction PDFs are all drawn from a given intrinsic color distribution in a control field. Hence, any objects which are not represented in the control field, for example YSOs in star-forming regions, will also use the given intrinsic feature set (e.g. main-sequence stellar colors or galaxy colors) for the extinction estimate. For a correct de-reddening of YSOs it is therefore necessary to use intrinsic features of such sources instead of typical main sequence or galaxy features. This issue is mitigated to some degree because typical intrinsic NIR YSO colors for classical T Tauri stars \citep[e.g.][]{meyer97} are found to be very similar to intrinsic galaxy colors. Furthermore, for extinction mapping of star-forming molecular clouds, YSOs should be removed beforehand from the input source list as they do not sample the full cloud column-density.

% -----------------------------------------------------------------------------
\subsubsection{Discretization}
\label{sec:discretization}

If a single value for the extinction is desired, it is possible to calculate the expected value (maximum probability, or any other meaningful descriptor) of this distribution. The uncertainty in the calculated discrete extinction can be estimated with the population variance of the distribution:
\begin{eqnarray}
    \mu_{f}          &=&  \int \! x \, f(x) \, \mathrm{d}x \\
    \mathrm{Var}_{f} &=& \int \! x^2 \, f(x) \, \mathrm{d}x - \mu^2.
\end{eqnarray}
Here $\mu$ refers to the expected value of a probability density function $f(x)$. In the case of a one-dimensional Gaussian Mixture Model, the mean and variance of the mixture can be written in terms of the means, variances, and weights of its components.
\begin{eqnarray}
    \mu_{\mathrm{mixture}}          &=&  \sum_i \mu_i w_i \label{eqn:ev} \\
    \mathrm{Var}_{\mathrm{mixture}} &=& \sum_i w_i \sigma_i^2 + \sum_i w_i \mu_i^2 - \left( \sum_i w_i \mu_i \right)^2 \label{eqn:var},
\end{eqnarray}
where $\mu_i$ refers to the mean of the $i$-th gaussian component, the $w_i$ are their weights with $\sum w_i = 1$, and $\sigma_i^2$ are the variances.

% -----------------------------------------------------------------------------
\subsection{Creating smooth extinction maps}

\pnicers derives PDFs for single sources which describe the line-of-sight extinction. To create extinction maps it is therefore necessary to construct the smooth column-density distribution from these irregularly spaced samples. If a single value of the extinction is derived from the PDFs (e.g. the expected value or the maximum probability) one can employ the well tested approach of the \nicers method \citep[][]{lombardi2001}. \pnicers also comes with built-in fully automatic solutions to create extinction maps with valid world coordinate system projections.

The \pnicers software package offers a variety of estimators for the smoothing process including nearest-neighbor, gaussian, and simple average or median methods. We note here that extinction maps constructed from discrete measurements can suffer from foreground contamination and unresolved cloud substructure which may introduce a bias in the column-density measurement. For more distant clouds foreground stars may represent the majority of detected sources in a pixel of the extinction map. These issues are described in more detail in \citet{lombardi2009} where the authors also introduce a new technique, \nicest, to minimize this bias. For discretized extinction measurements \pnicers optionally also includes this method. Constructing smooth column-density maps using the full probabilistic description of extinction via GMMs will be the subject of a follow-up study.

% -----------------------------------------------------------------------------
% -----------------------------------------------------------------------------
\section{Method validation}
\label{sec:validation}

In order to evaluate the new method we compare results from \pnicers to \nicers by directly deriving extinction and associated errors via the expected value and the population variance of the PDFs. In this section we

\begin{itemize}

    \item[a)] analyze results when both algorithms are applied to the VISION Orion~A data.

    \item[b)] investigate the intrinsic color distribution as measured in the VISION control field and discuss the bias of comparing different observed populations when the science field is extincted. This part will also highlight the effects of using increasing numbers of parameters.

    \item[c)] compare wide-field extinction maps calculated with both techniques from 2MASS data.

\end{itemize}
In addition we also examine the software performance and describe where interested users can access the source code and potentially even contribute to the development. We will not discuss the relation of the estimated errors for both methods, as they are derived in different ways and are therefore not comparable.

% -----------------------------------------------------------------------------
\subsection{Applying \nicers and \pnicers to real data}
\label{sec:real_data}

\begin{figure}
	\centering
    \resizebox{\hsize}{!}{\includegraphics[]{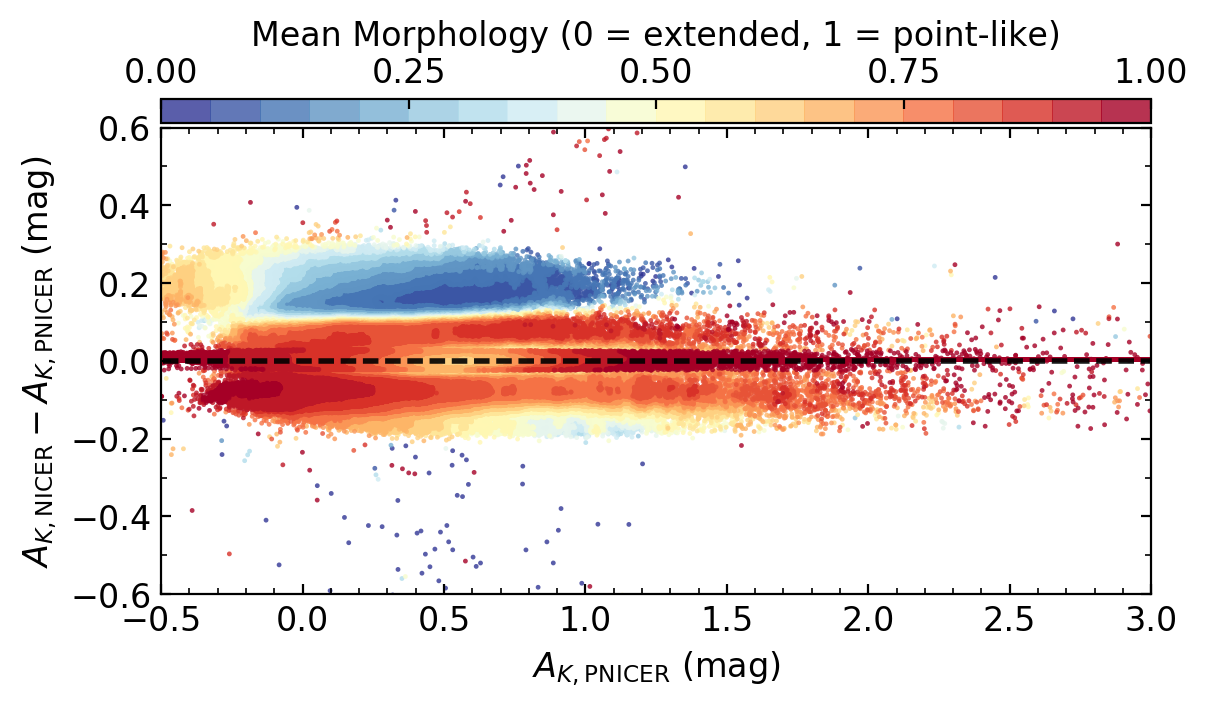}}
	\caption{Direct comparison of \pnicers and \nicers extinction estimates when applied to the NIR VISION observations. The mean morphology was calculated with the SExtractor morphology class of the source catalog in a $0.05 \times 0.05$ large box in this parameter space. As expected, \pnicers delivers smaller extinction estimates for galaxies.}
	\label{img:pnicer_vs_nicer_morphology}
\end{figure}

In a first evaluation of the \pnicers method, we applied the algorithm together with \nicers to the photometric color data of the VISION NIR observations (no magnitudes in parameter space, only $J,H,K_S$). For \pnicers we use the expected value of the PDFs and their population variances to directly calculate extinction and errors (see equations \ref{eqn:ev} and \ref{eqn:var}). Taking the expected value of the PDFs makes the methods directly comparable since \nicers relies on a similar method (see Sect. \ref{sec:motivation} for details). The difference in the derived color-excesses is shown in Fig.~\ref{img:pnicer_vs_nicer_morphology} where the data are color-coded by source morphology. At first glance, the distribution appears to be bimodal and we clearly see that for galaxies (morphology $\approx$ 0) on average a much larger difference between the methods is seen when compared to point-like sources. As expected \pnicers delivers smaller extinction towards galaxies when compared to \nicers, because the latter method places the mean intrinsic color between the galaxy and M-sequence locus (compare Fig.~\ref{img:control_covar}). At closer examination, however, we observe a more complex structure: there is a distinguished distribution at $A_{K, \mathrm{NICER}} - A_{K, \mathrm{PNICER}} = 0$ which is caused by sources having only measurements in two photometric bands. In this case (\pnicers has only access to colors) the results of \nicers and \pnicers are identical. Additionally, there is also an enhancement of sources towards negative values in $A_{K, \mathrm{NICER}} - A_{K, \mathrm{PNICER}}$. This is caused by sources which are de-reddened beyond the assumed mean of \nicers, which, in this case, are mostly stellar sources, since, as already mentioned above, the mean of the intrinsic color distribution is found between galaxy and stellar loci.

\begin{figure*}
	\centering
    \resizebox{\hsize}{!}{\includegraphics[]{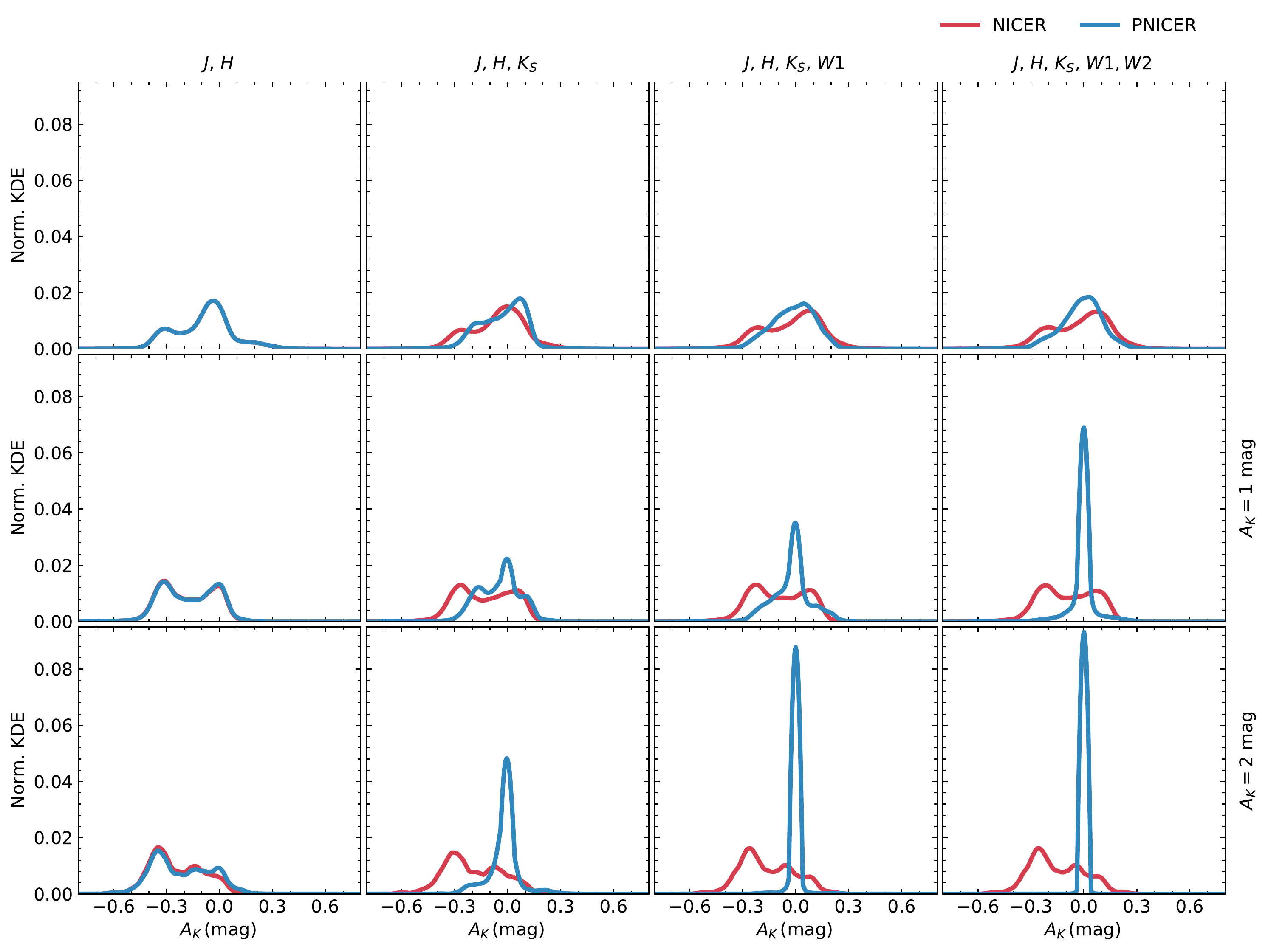}}
	\caption{Kernel densities for the derived line-of-sight extinction values for both \pnicers (blue) and \nicers (red) when applying the algorithms to the VISION control field itself. In this case and since the entire control field is extinction-free, the calculated values for should be close to zero. From left to right the number of features is incrementally increased, while from top to bottom we apply magnitude cuts to the ``science'' field. The magnitude cuts simulate the effects of having an extincted science field, resulting in different populations when compared to the extinction-free control field (e.g. galaxies will be shifted beyond the survey sensitivity limit). In the first row we use all data, while the second and third row simulate the effect of 1 and 2 mag extinction in the $K_S$ band (i.e. $K_S < 18$ mag and $< 17$ mag, respectively for our completeness of $K_S = 19$ mag). Clearly, \pnicers performs better for increasingly different populations and number of available features.}
	\label{img:pnicer_vs_nicer}
\end{figure*}

% -----------------------------------------------------------------------------
\subsection{The intrinsic color distribution and population bias}
\label{sec:intrinsic_colors}

For the \pnicers and \nicers methods an extinction-free control field typically observed at similar galactic latitudes as the science field, by assumption, holds the information for intrinsic features. Therefore, when the techniques are applied to the control field itself, ideally the measured extinctions should be close to 0. We test this hypothesis under a variety of circumstance: we (a) vary the number of available parameters and (b) additionally apply magnitude cuts to the ``science'' field (for this test the control field itself) while always using the same (only sensitivity limited) control field data. The idea behind point (b) is to simulate the effect of extinction. While in an extincted field intrinsically faint sources will be shifted beyond the photometric sensitivity limit (i.e. fewer faint stars and galaxies will be observed), the control field does not suffer from these effects. Furthermore, to increase the dimensionality for these tests we combine the NIR VISION control field photometry with the first two bands of the ALLWISE source catalog. \nicer, as usual, is restricted to color information, but to highlight additional differences we allow \pnicers to construct the multidimensional feature space from both color and magnitude information simultaneously. Thus, \nicers has access to up to four dimensions (colors only), while \pnicers has access to nine dimensions (five magnitudes and four colors) at most.

Figure~\ref{img:pnicer_vs_nicer} displays the results when applying both the \nicers and \pnicers algorithms to the VISION control field itself. All panels show kernel densities (``histograms'', bandwidth = 0.04 mag) for the distributions of the derived extinction (for \pnicers again the expected value of the PDF), where the blue lines refer to \pnicers results and the red lines to \nicer. The separate columns in the figure refer to different parameter combinations with increasing dimensionality from left to right. In the first column, the analysis is restricted to $J$ and $H$ only, while in the last column we show the results when using $J$, $H$, $K_S$, $W1$, and $W2$ photometry. The different rows in this plot matrix refer to different magnitude cuts for the science field (the control field remains untouched). Applying magnitude cuts to the data simulates the effects of extinction on the observed population. Consider an extinction of $A_K = 1$ mag and our sensitivity limit at $K_S = 19$ mag. In this case all sources which have intrinsic apparent magnitudes between $K_S = 18$ and 19 mag will shift beyond the detection limit, creating a different observed population with a different feature (e.g. color) distribution. Therefore, limiting our data to $K_S < 18$ mag (and the other bands according to Table \ref{tab:extinction_law}) has the same effect as 1 mag of extinction in this band. In Fig.~\ref{img:pnicer_vs_nicer} the first row does not apply magnitude cuts, while the second and third row simulate the effects of having an extinction of $A_K = 1$ and 2 mag, respectively. We note here that the calculated extinction for individual sources can become negative when the estimated intrinsic color is ``redder'' than the observed color for the investigated source. This is possible for unextincted sources and especially affects early-type stars. As a consequence, the kernel densities extend both to the positive and negative side in this analysis. 

\begin{figure*}
	\centering
    \resizebox{\hsize}{!}{\includegraphics[]{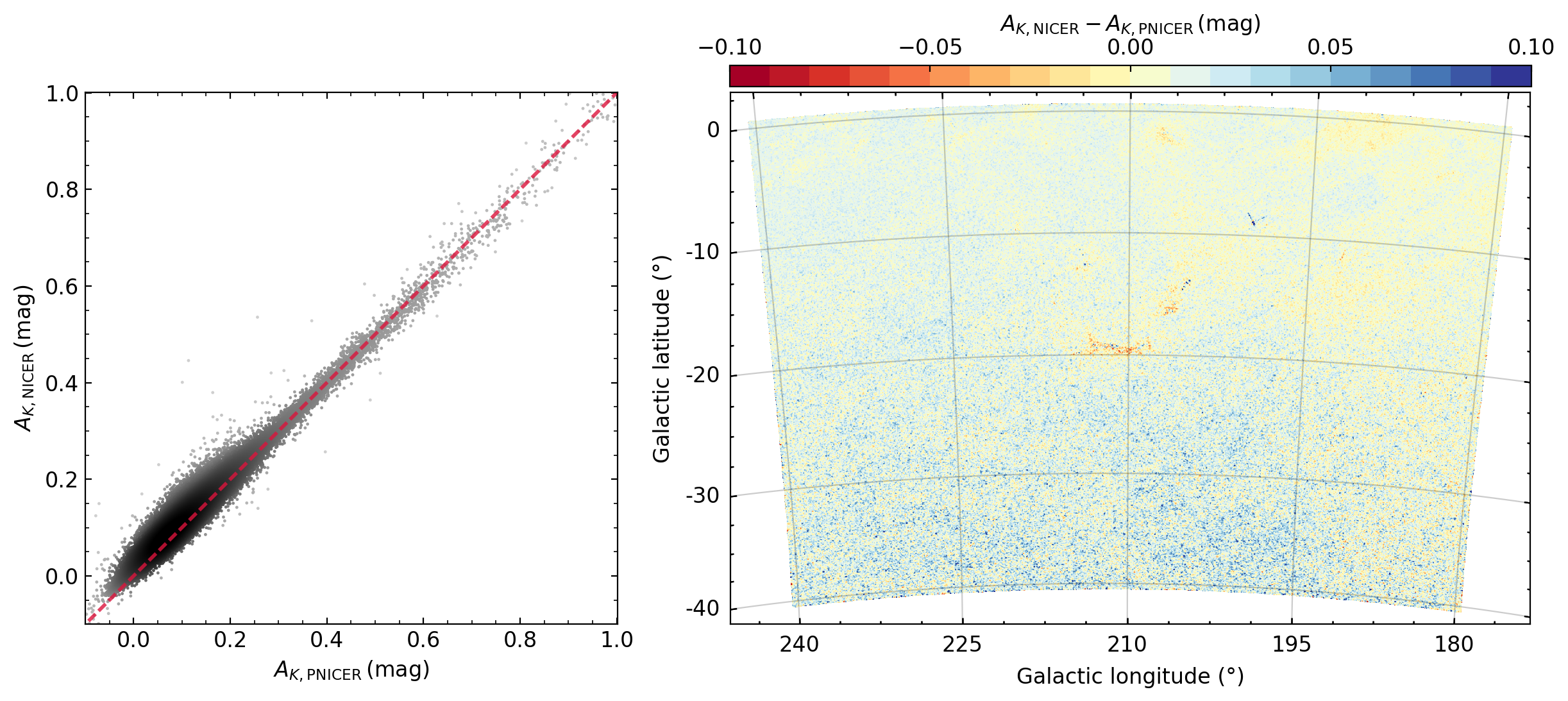}}
	\caption{Comparison of \pnicers and \nicers when applied to 2MASS data. Both panels shows a pixel-by-pixel comparison of extinction maps (5~arcmin resolution) calculated with both methods from the same data. As expected, within the statistical error the results are identical. Nevertheless we observe a small systematic shift towards larger \nicers extinctions in low column-density regions. The relatively large deviation towards the center of the map is attributed to the Orion~A and B molecular clouds where extinction estimates have even larger variance.}
	\label{img:nicer_vs_pnicer_wide}
\end{figure*}

For only two features ($J$ and $H$; leftmost column) we obtain very similar results for both methods across all magnitude cuts. However, for increasingly strict magnitude cuts (more extinction) we can see that the peak in the distribution is shifted towards negative values. This is expected, since in this scenario only stellar sources remain in the science field while the mean color of the full control field is calculated from data including galaxies. In this case both methods are biased since there is not enough information to break degeneracies in the intrinsic feature space. The resulting extinction distributions are overall almost identical, but due to allowing also magnitude information, the \pnicers extinction distribution can appear slightly different. Here we note again that when considering only one color without magnitude information the results from \pnicers and \nicers are identical. 

When increasing the number of available features to three ($J,H,K_S$, second column) we already observe significant differences. While \nicers still suffers from the bias of different populations in the science and control field (the peak is again shifted to negative $A_K$), \pnicers starts to perform systematically better. For a simulated extinction of $A_K = 1$ and 2 mag (i.e. $K_S < 18$ and 17 mag) the increased dimensionality helps to break degeneracies in the intrinsic feature space and the extinction distribution shows a prominent peak at $A_K = 0$ mag. This effect is even more pronounced when including four or five parameters (third and fourth column) where \pnicers overall shows systematically better results than \nicer. Especially in the case of largely different populations in the science and control field \pnicers manages to overcome this issue and delivers far better results. We note here that for the case of similar observed populations (top row), \pnicers is only marginally better than \nicers ($\sim$20 -- 30\% narrower distribution width when using five features), but we expect that the remaining degeneracy in intrinsic colors could be better lifted by including one or two more suitable passbands (e.g. $Y$ at 1 $\mu$m). Especially including bands towards optical wavelengths would help in this case, since here the stellar sequences are typically more pronounced than in the NIR allowing \pnicers to better separate object types. The practical limit in the number of features here depends on the sensitivity of the observations in the given passbands (higher extinction towards bluer bands) and the sampling of the control field feature space (more dimensions require more sources for accurate sampling). We conclude that \nicers is biased in cases where the populations in the science and control field are different, i.e. in regions with extinction. \pnicers on the other hand starts to break the degeneracy in intrinsic colors when having access to more than one color-feature and performs even better when including more parameters.

% -----------------------------------------------------------------------------
\subsection{Extinction maps}
\label{sec:extinction_maps}

We also validated the new method's functionality by comparing wide-field extinction maps created with \nicers and \pnicers based on 2MASS data only. As in the tests above, we calculated the extinction towards each source as the expected value of the associated PDF for \pnicer. Since the data are restricted to the three NIR bands $J$, $H$, and $K_S$ and only include a negligible number of galaxies (if any) the color distributions are relatively narrow and we expect very similar results without considerable improvement in the quality of the extinction map. Nevertheless, this test should demonstrate that under these simple circumstances \pnicers works equally well as \nicers.

For this purpose we created extinction maps with a resolution of 5 arcmin for approximately the same region as the maps in \citet[][]{lombardi2011} who studied a $\sim$40 $\times$ 40 deg$^2$ field including the Orion, Monoceros R2, Rosette, and Canis Major star forming regions. The control field in this case was chosen as a $2\times2$ deg$^2$ wide sub-region centered on $l=233.3$, $b=-19.4$, the position of the VISION control field. We note here that for such a large region it would be more appropriate to use multiple control fields located at different galactic latitudes to accurately sample the galactic stellar population. This application, however, only serves as a demonstrator and to verify the method. Therefore variations in the field population can safely be neglected.

The results of this test are displayed in Fig.~\ref{img:nicer_vs_pnicer_wide} where we show pixel-by-pixel comparisons for both maps. The results can be considered equal within the noise properties of the data, but nevertheless we see a small systematic trend of \nicers giving slightly larger extinction values for low column-density regions when compared to \pnicer. The Orion~A molecular cloud (approximately located at the center of the map) is also well visible in this comparison in the right-hand side panel of Fig.~\ref{img:nicer_vs_pnicer_wide}. In this case we do not interpret this behaviour as systematic differences between the methods since the noise levels also significantly increase in this region due to high column-densities and fewer available background sources. Nevertheless, to some extent, this may be caused by the above discussed population bias. Hence, we conclude that \pnicers and \nicers work equally well for cases were only the typical NIR bands are considered and galaxies do not contaminate the intrinsic color space.

\begin{figure*}
	\centering
	\resizebox{\hsize}{!}{\includegraphics[]{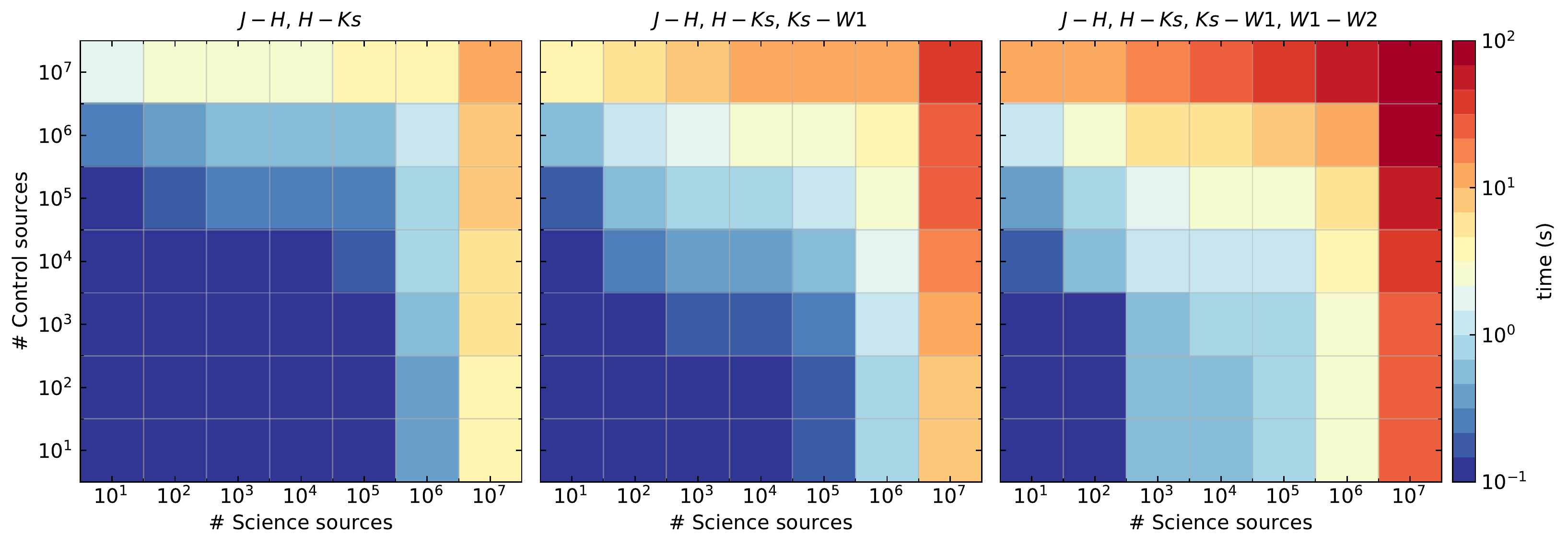}}
	\caption[]{\pnicers performance results. For the tests displayed in the left-most panel two features were used ($J-H$, $H-Ks$), for the panel in the center three features, and for the right-most results four features. The colors of each matrix element display the code execution time for a given setup where the number of sources in the science field and the number of sources in the control field are incrementally increased. While for typical applications for a few thousand to a few ten thousand sources the de-reddening process finishes in a fraction of a second, only for extreme (and probably rare) use cases ($10^7$ sources) we measure execution times longer than a minute.}
	\label{img:performance}
\end{figure*}

% -----------------------------------------------------------------------------
\subsection{Performance}
\label{sec:perfomance}

We evaluated the performance of our \pnicers implementation by generating several test scenarios with variable numbers of science and control field sources. All performance tests have been conducted on a machine with a 4~GHz CPU (Intel Core\texttrademark i7-6700K) with 16 GB RAM. For all tests we used the VISION control field for the intrinsic feature distribution and applied randomly generated extinction to the sources (according to Table \ref{tab:extinction_law}) to simulate extincted science fields. From this data pool we then randomly drew sets of variable size by (a) varying the number of sources in the science field, (b) varying the number of control field sources, and  (c) increasing the number of available features.

The results of these performance tests are visualized in Fig.~\ref{img:performance}. The individual panels represent different numbers of features (two, three, and four from left to right). Each matrix element in the panels shows the results of a single \pnicers run. For example the bottom right elements show the run time results of using $10^7$ science field sources with a control field that only contains 10 sources\footnote{The minimum number of 20 sources along the reddening vector has been disabled for this test.}. The most extreme case in the top right uses $10^7$ sources in both the science and control fields. For small sample sizes ($\lesssim 10^4$), runtimes are mostly found well below 1~s across all feature combinations. For a VISION-type use case ($10^6$ science sources, $10^5$ control field sources and using only two colors) the execution time of \pnicers is 0.8~s. Only for very extreme cases of $10^7$ sources and more than two features, the de-reddening process requires more than a minute. Extrapolating these results to cases using a combination of more than 4 features, we recommend to keep the sample size below $10^6$. With these results we therefore conclude that the \pnicers performance is highly competitive and that the application of the method also remains practical in extreme cases.

% -----------------------------------------------------------------------------
\subsection{Software availability}
\label{sec:availability}

\pnicers is an open-source software and is accessible to all interested users. The Python package and source code are available at \url{https://github.com/smeingast/PNICER} where also the latest versions will be made available. Future upgrades may include advanced treatment of photometric errors, support for additional metrics beyond photometric measurements, and weighted fitting of the Gaussian Mixture Models as soon as the necessary libraries are updated or become available.

% -----------------------------------------------------------------------------
% -----------------------------------------------------------------------------
\section{Summary}
\label{sec:summary}

We have presented a new method, \pnicer, to derive extinction probability density functions for single sources in arbitrary numbers of dimensions. Our findings can be summarized as follows:

\begin{enumerate}

    \item The well-established \nicers method to calculate line-of-sight extinctions suffers from increasing variance in intrinsic color estimations for deep NIR observations when galaxies enter the color space. As a consequence, the color excess estimates have large statistical errors. For details see Fig.~\ref{img:control_covar}. Other methods do not implement satisfying solutions for this problem since they either rely on additional information (e.g. morphology) or are restricted to few dimensions and are computationally expensive.
    
    \item We introduce a new method, \pnicer, which uses unsupervised machine learning tools to calculate extinction towards single sources. To this end we fit the intrinsic feature distribution from an extinction-free control field with Gaussian Mixture Models. The resulting probability density distribution describes the probability of intrinsic features (e.g. colors) and therefore also extinction. From these distributions the extinction can be estimated with the expected value (or maximum probability), its uncertainty with the PDF variance. Details on the method are visualized in Fig.~\ref{img:pnicer_method}.
    
    \item \pnicers is entirely data-driven and does not require prior information of source characteristics or the column-density distribution. The intrinsic feature probabilities are automatically constructed from the control field data with unsupervised algorithms. 
    
    \item \pnicers works in arbitrary numbers of dimensions and features (e.g. magnitudes or colors) can be combined in any way as long as the corresponding extinction law is known.
    
    \item We investigated the effects of the intrinsic color distribution and compared the \pnicers and \nicers performance when the observed populations in the extincted science and the extinction-free control field are different (i.e. in regions of significant extinction). We find that \nicers is biased when different populations are observed and that \pnicers performs significantly better in these cases. To break degeneracies in the intrinsic feature space with \pnicer, more than one parameter is required (e.g. two colors). For details see Fig.~\ref{img:pnicer_vs_nicer}.
    
    \item Using 2MASS data (three NIR bands, no galaxies) \pnicers reproduces the \nicers extinction mapping results within the statistical errors (Fig.~\ref{img:nicer_vs_pnicer_wide}).
    
    \item The \pnicers software is entirely written in Python and is publicly available at \url{https://github.com/smeingast/PNICER}. It includes simple interfaces to apply either the \pnicers or the \nicers method to real data and subsequently construct extinction maps. Furthermore, the algorithm remains computationally competitive for large ensembles, calculating extinction PDFs for millions of sources in a matter of seconds.

\end{enumerate}

% ----------------------------------------------------------------------
% ----------------------------------------------------------------------
\begin{acknowledgements}
Stefan Meingast is a recipient of a DOC Fellowship of the Austrian Academy of Sciences at the Institute for Astrophysics, University of Vienna.
We thank Kai Polsterer for the helpful discussions on Machine Learning and his valuable input regarding the methods presented in this publication.
We also thank the anonymous referee for useful comments that helped to improve the quality of this publication.
This research made use of Astropy, a community-developed core Python package for Astronomy \citep{astropy}.
\end{acknowledgements}

% ----------------------------------------------------------------------
% ----------------------------------------------------------------------
\bibliography{bibliography}

% ----------------------------------------------------------------------
% ----------------------------------------------------------------------
\begin{appendix}

% ----------------------------------------------------------------------
% ----------------------------------------------------------------------
\section{Software dependencies, structure, and availability}
\label{sec:software}

One of the main method and software design goals for \pnicers was to make it accessible and usable for as many people as possible. For this reason we have kept the number of required dependencies at a minimum and we additionally set a high value for the computational performance. The entire software is written in Python (\url{https://www.python.org}) and its main dependencies are NumPy \citep{numpy} and SciPy \citep{scipy} for numerical calculations, Astropy \citep{astropy} for I/O and world coordinate system support, and scikit-learn \citep{scikit-learn} for machine learning tools. In addition, the plotting methods make use of matplotlib \citep{matplotlib} and the Astropy affiliated wcsaxes package (\url{https://github.com/astrofrog/wcsaxes}). All of these packages are easily accessible through the Python Package Index (\url{https://pypi.python.org/pypi}).

For high performance some functions in \pnicers have have been parallelized to offer even better results on modern machines. The Python parallelization oftentimes works with straight-forward code-blocks in Python, however, the multiprocessing library of our choice is (at the time of writing this manuscript) not compatible between Unix-based operating systems and Windows. Therefore, at the moment, \pnicers can only be used on Unix machines. All essential tests have been successfully performed under macOS 10.12 and Ubuntu 14.04 LTS and we do not foresee any major compatibility issues with future operating system or Python versions. Using our software implementation and its functions only requires to instantiate e.g. photometric data as \textit{ApparentColors} or \textit{ApparentMagnitudes} objects. Once the instance is created running \pnicers or \nicers only requires a single line of code. Subsequently running the discretization and creating an extinction map from the extinction estimates also only require an additional single line of code. All customization and options for running the software are implemented as keyword arguments in the \pnicers or \nicers call. 

Future versions may offer enhanced construction of extinction maps and support for additional feature metrics. In its current form the software only allows photometric data to be instantiated, but in principle any metric can be used and combined with other parameters as long as the extinction can be described in the same way as for magnitudes or colors. Running the algorithms returns \textit{ContinuousExtinction} objects from which discretized \textit{DiscreteExtinction} objects and subsequently extinction maps can be created. In its simplest form a typical \pnicers session may look like the following example.

\begin{lstlisting}[language=Python]
# Import PNICER class for apparent colors
from pnicer import ApparentColors as AC

# Instantiate Color objects from observed data
sci = AC(colors=mag_sci, errors=err_sci, ...)
con = AC(colors=mag_con, errors=err_con, ...)

# Calculate extinction PDFs
pdf = sci.pnicer(control=con)
 
# Discretize PDFs
extinction = pdf.discretize()

# Construct the extinction map
emap = extinction.build_map(bandwidth=0.1)

# Save the map with WCS projection
emap.save_fits(path="/path/to/map.fits")
\end{lstlisting}

\end{appendix}
\newpage

% ----------------------------------------------------------------------
% ----------------------------------------------------------------------
\end{document}